\documentclass[a4paper,10pt,twoside]{cpc-hepnp}
\usepackage{CJK,upgreek,fancyhdr}
\usepackage{appendix}
\usepackage{revsymb}
\usepackage{multicol}
\usepackage{multirow}
\usepackage{graphicx}
\usepackage{booktabs}
\usepackage{amssymb,bm,mathrsfs,bbm,amscd}
\usepackage[tbtags]{amsmath}
\usepackage{lastpage}
\usepackage[colorlinks=true, urlcolor=blue, linkcolor=red]{hyperref}

\newcommand{\Correction}[1]{\textcolor{black}{#1}}

\begin{document}
\begin{CJK*}{GB}{gbsn}

\fancyhead[c]{\small Chinese Physics C~~~Vol. xx, No. x (201x) xxxxxx}
\fancyfoot[C]{\small 010201-\thepage}

\footnotetext[0]{Received \today}

\title{Measurement of yields and angular distributions of $\gamma$-quanta from the interaction of $14.1$~MeV neutrons with oxygen, phosphorus and sulfur\thanks{This work was supported by a RSCF grant No.23-12-00239}}

\author{%
      Grozdanov D.N.$^{1,2;1)}$\email{dimitar@nf.jinr.ru}%
\quad Fedorov N.A.$^{1,3}$
\quad Dabylova S.B.$^{1,4}$
\quad Kopatch Yu.N.$^{1}$
\quad Ruskov I.N.$^{2}$\\
\quad Skoy V.R.$^{1}$
\quad Tretyakova~T.Yu.$^{1,3,5}$
\quad Hramco C.$^{1,6}$
\quad Kharlamov P.I.$^{1,5}$
\quad Pampushik G.V.$^{3}$\\
\quad Filonchik P.G.$^{1,7}$
\quad Andreev A.V.$^{1,3}$
}

\maketitle

\address{%
$^1${Joint Institute for Nuclear Research (JINR), Dubna, Russia}\\
$^2${Institute for Nuclear Research and Nuclear Energy, Bulgarian Academy of Sciences, Sofia, Bulgaria}\\
$^3${Faculty of Physics, Lomonosov Moscow State University, Moscow, Russia}\\
$^4${L.N.Gumilyov Eurasian National University, Astana, Kazakhstan}\\
$^5${Skobeltsyn Institute of Nuclear Physics, Moscow State University, Moscow, Russia}\\
$^6${Institute of Chemistry of Moldova, Chisinau, Republic of Moldova}\\
$^7${Moscow Institute of Physics and Technology, Moscow, Russia}
}

\begin{abstract}
A study of the inelastic scattering of neutrons with an energy of $14.1$~MeV on the nuclei of oxygen, phosphorus and sulfur was carried out at the TANGRA facility at JINR (Dubna). The purpose of the experiment was to refine existing and obtain new data on the yields and angular distributions of $\gamma$-quanta emitted by the studied nuclei as a result of neutron-induced nuclear reactions using the tagged neutron method. Two types of detector systems were used to register $\gamma$-quanta. The $\gamma$-ray yields were measured using a high-purity germanium (HPGe) detector. The angular distributions of $\gamma$-rays were obtained using a system of 18 scintillation detectors based on bismuth germanite Bi$_{4}$Ge$_{3}$O$_{12}$ (BGO) located around the sample. As a result of the studies carried out, the yields of two transitions for the reaction of tagged neutrons with $^{16}$O, nine transitions for the reaction with $^{31}$P, and nine transitions for the reaction with $^{32}$S were measured for the first time. The angular anisotropy of the $\gamma$-radiation accompanying the inelastic scattering of neutrons with an energy of $14.1$~MeV on $^{31}$P nuclei was also measured for the first time.
\end{abstract}



\footnotetext[0]{\hspace*{-3mm}\raisebox{0.3ex}{$\scriptstyle\copyright$}2013
Chinese Physical Society and the Institute of High Energy Physics
of the Chinese Academy of Sciences and the Institute
of Modern Physics of the Chinese Academy of Sciences and IOP Publishing Ltd}%

\begin{multicols}{2}

\section{Introduction}

Information about neutron-nucleus interactions is extremely important for both fundamental and applied physics. Neutral electric charge of the neutron makes it a unique probe for studying the structure of matter both at the molecular and nuclear levels, as well as for revealing the nature of nuclear forces. Data on neutron-nuclear reactions are also necessary for designing nuclear power plants, as well as for modeling various devices and objects that interact with neutron radiation in one way or another. For example, neutrons are widely used in inspection complexes and installations for neutron logging, non-destructive elemental analysis of matter, nuclear medicine, production of radio-pharmaceuticals, etc. An indicator of the relevance of studying the characteristics of neutron-nucleus interactions can be the fact that the NEA Nuclear Data High Priority Request List (HPRL) \cite{NEA} mostly consists of queries directly related to neutron-nuclear reactions. The most important task facing the nuclear power industry is the creation of Generation IV fast neutron reactors, which requires both high safety and reliability of all technological processes, and economic efficiency. The design and construction of new reactors imposes serious requirements on the quality of neutron-nuclear data for the largest possible number of nuclides in a wide range of neutron energies, especially for fast neutrons.

Oxygen is one of the most abundant elements on earth. In addition to water, it is part of a large number of chemical compounds -- mainly oxides, which leads to the need to focus on its study \cite{NEA}. Phosphorus and sulfur are also widespread. For example, they are components of apatite, a valuable raw material for the production of mineral fertilizers. Timely control of the elemental composition of apatite ores is important for the correct passage of technological processes, and installations for the rapid elemental analysis of this raw material have already been created \cite{Bolshakov2020}. Sulfur, in turn, is part of a large number of products of the chemical industry, in particular rubbers and explosives. It is a pollutant of oil products, significantly worsening their properties, and therefore the creation of a technique that allows measuring its content is of great importance.

Our research is dedicated to studying the characteristics of $\gamma$ radiation arising from the interaction of fast neutrons with atomic nuclei. On the one hand, $\gamma$-quanta are a marker of the presence of the corresponding nuclide in the medium, and on the other hand, this radiation is an important component of inelastic neutron scattering and its characteristics are necessary for complete modeling of neutron-nuclear interaction processes. The experiments were carried out at Frank Laboratory of Neutron Physics (JINR, Dubna) in the framework of the TANGRA (TAgged Neutrons and Gamma RAys) project to study the inelastic scattering of neutrons with an energy of 14.1 MeV with atomic nuclei using tagged neutron method (TNM) \cite{2015a,2015b}.

The TNM is based on registration of the $3.5$~MeV $\alpha$-particle from the reaction
\begin{equation}
\label{eq:d_t_reaction}
d + t \rightarrow n + \alpha.
\end{equation}

According to the reaction's kinematics, the neutron is emitted in the direction opposite to that of the $\alpha$-particle (in the center-of-mass frame). Thus, knowing the direction of emission of the $\alpha$-particle, it is possible to determine the direction of motion of the neutron. This way, it is possible to form a beam of "tagged" neutrons with a well-defined energy and flux. The $\alpha$-particles are registered in coincidence with the pulses from the characteristic nuclear $\gamma$-radiation, emitted from excited products of neutron-induced reactions on the nuclei $A$ in the sample:
\begin{equation}
\label{eq:gamma_reaction}
A(n,x)B^* \xrightarrow{\gamma} B.
\end{equation}
Using this time-correlated associated particle technique (TC-APT), which significantly reduces the effect of background radiation on the quality of experimental data, a multi-pixel $\alpha$-particle detector and a set of detectors for products of neutron-induced nuclear reactions, it is possible to carry out elemental and/or isotopic analysis (visualization, radiography, tomography) of any substance and/or object, to determine the differential, and total cross sections of neutron-induced nuclear reactions, as well as the energy and angular distributions of these reactions' products with higher accuracy and reliability.

Continuing the experimental study of the inelastic scattering of neutrons with an energy of 14 MeV by some light and medium nuclei \cite{Grozdanov2018,Fedorov2021,Fedorov2020,Grozdanov2020,Dabylova2021} is carried out within the framework of the TANGRA project. Here we present the results on the yields and angular distributions of $\gamma$-quanta emitted in the $(n, X\gamma)$ type reactions on oxygen, phosphorus, and sulfur nuclei, where $X = n', p, d, \alpha$.

\section{Experimental setups}

A portable neutron generator ING-27 manufactured by VNIIA (Moscow, Russia) was used as a neutron source. It is a small-size device based on sealed gas-filled neutron tube with a built-in multi-pixel $\alpha$-particle detector that could produce maximal neutron flux of $5 \cdot 10^7$~s$^{-1}$.
Reaction (\ref{eq:d_t_reaction}) is induced by a continuous beam of deuterons with a kinetic energy of $80 - 100$~keV focused on a titanium tritide target.

The position-sensitive $\alpha$-particle detector placed in the ING-27 neutron tube at a distance of 100 mm from the target consists of 16 (8 + 8) mutually perpendicular semiconductor strips, each 6 mm wide and 55 mm long. The overlapping strips form 64 (8x8) pixels with a pixel size of 6x6 mm.
The count rate of each pixel of the $\alpha$-detector uniquely determines the associated beam of tagged neutrons. 
This way, depending on neutron source-to-target solid angle, one can use up-to 64 tagged neutron beams of well defined energy, direction and flux.

The TANGRA setup can be used with multifunctional detector systems of various configurations. To study the properties of nuclear reactions that occur when samples are irradiated with 14-MeV neutrons, two detector systems were used: the "Romasha" system with 18 BGO (Bi$_4$Ge$_3$O$_{12}$) scintillation $\gamma$-ray h-purity germanium  (HPGe) $\gamma$-ray spectrometer. \Correction{The energy resolution for the BGO detectors was about 110 keV at 1.33 MeV, and for the HPGe detector -- 1.9 keV.}

The layout of the TANGRA setup for $\gamma$-quanta yields measurement using a HPGe $\gamma$-ray detector is shown in Fig.~\ref{fig:hpge}. The HPGe crystal has a diameter of $57.5$~mm and a thickness of $66.6$~mm. The detector was located at the minimum possible distance from the sample and protected from the direct neutrons emitted from the generator by a 172~mm thick lead shielding. With this configuration, the energy spectra of $\gamma$-rays were measured with high resolution. 

 Fig.~\ref{fig:bgo} presents the layout of the TANGRA setup for  the $\gamma$-radiation angular distribution measurement using the "Romasha" detector system. It consisted of 18 BGO gamma scintillation detectors arranged in 14 degree increments in a circle with a radius of 75 cm, and a sample in its center.
 
The samples used in this experiment were rectangular boxes made of 0.1 mm thick aluminum foil filled with investigated substances. The bulk density of each sample was determined as a ratio of net weight to volume. The main characteristics of the samples are shown in Table~\ref{table:samples}.

\begin{center}
\includegraphics[width=75mm]{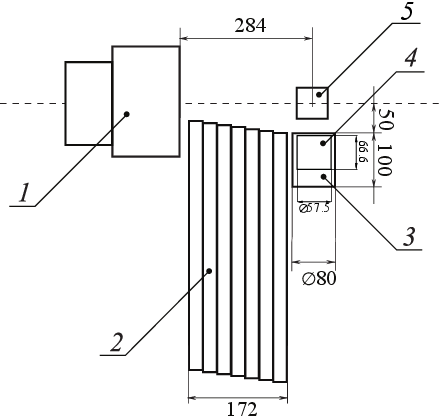}
\figcaption{Scheme of the TANGRA setup with the HPGe detector in the reaction plane: 1 -- neutron generator ING-27, 2 -- lead shielding, 3 -- housing of the HPGe detector, 4 -- HPGe crystal, 5 -- sample. The axis of the experimental setup is indicated by a horizontal dashed line. The tritium-enriched target is marked with an asterisk. All dimensions are given in mm.}
\label{fig:hpge}
\end{center}

\begin{center}
\includegraphics[width=75mm]{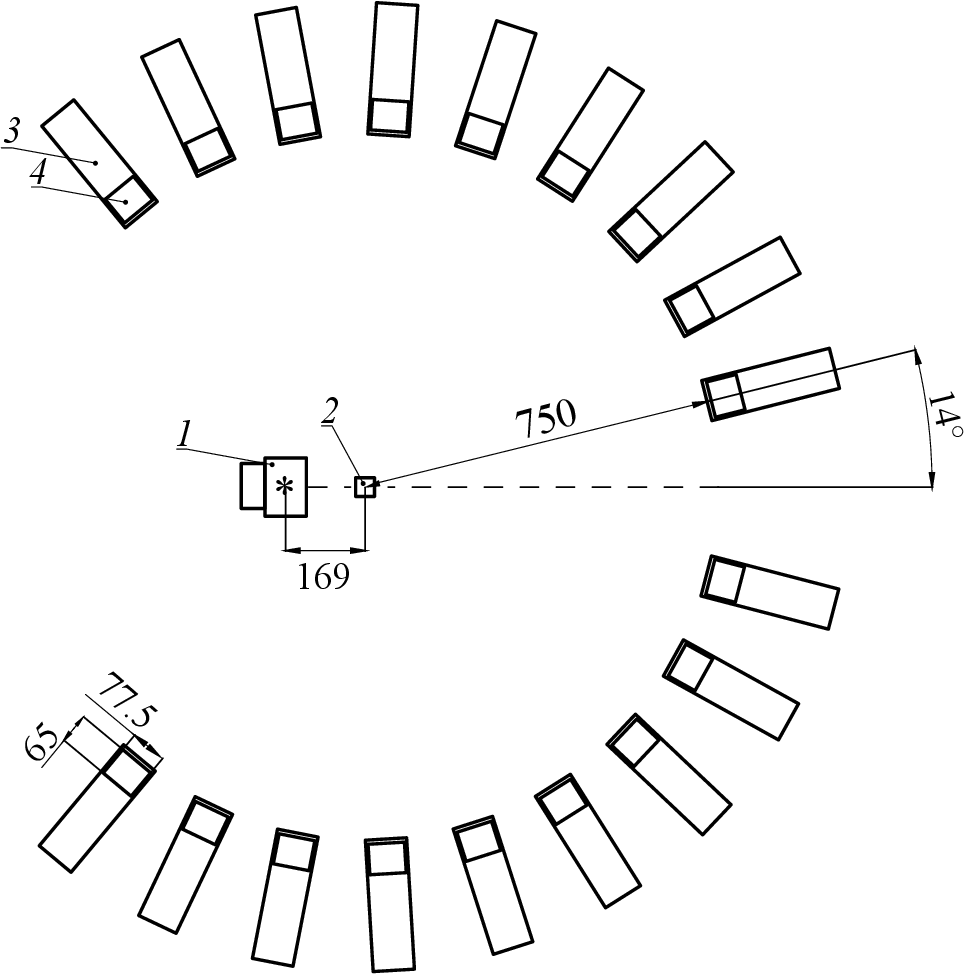}
\figcaption{\label{fig:bgo}Scheme of the TANGRA setup with BGO detectors in the reaction plane : 1 -- portable neutron generator ING-27, 2 -- sample in the center of Romasha $\gamma$-ray registration system, 3 -- BGO detector housing, 4 -- BGO crystall. The symmetry axis of the experimental setup is indicated by a horizontal dashed line. The tritium-enriched target is marked with an asterisk. All dimensions are given in mm.}
\end{center}

To collect and digitize the analog signals from the systems, two computerized ADCs were used: ADCM-32 for the "Romasha" detector system with BGO detectors and CRS-6/16 for experiments with the HPGe detector. Both devices were developed at JINR. The characteristics of the digitizers are shown in Table~\ref{table:adc}. The data acquisition system was controlled by the "Romana" computer program written by us using the CERN's ROOT \cite{ROOT}  framework. The list-mode data were saved on the computer hard drive for further off-line analysis using a number of ROOT-based programs.

The profiles of the tagged neutron beams were measured prior to the experiment using a position-sensitive silicon detector of charged particles which could be used for detecting fast neutrons via $^{28}$Si($n,\alpha$) reaction \cite{ZAMYATIN201846}. 
This information was used to optimize the sample dimensions and correct the position of the tagged neutron beams incident on it. \Correction{The horizontal dimension of the target (6 $\times$ 6 cm$^2$) were optimized to achieve $\gamma$-absorption not exceeding 20\% for considered $\gamma$-lines. Thus, only 4 neutron beams corresponding to 4 innermost vertical strips of the $\alpha$-detector covered the target.}

\end{multicols}

\begin{center}
\tabcaption{\label{table:samples} Sample characteristics}
\footnotesize
\begin{tabular}{ c  c  c  c  c c }
\toprule 
Sample & Bulk density (g/cm$^{3}$)  & Size (cm$^{3}$) & Mass (g) & Isotopic composition & Purity \\ 
\hline
P$_2$O$_5$ & 0.903 & 6 $\times$ 6 $\times$ 14 & 455 & $^{31}$P -- 100\%, $^{16}$O -- 99.757\%, & \textgreater 99\%  \\
&  &  &  &  $^{17}$O -- 0.038\%, $^{18}$O -- 0.25\% &  \\

\hline
S & 1.244 & 6 $\times$ 6 $\times$ 14 & 627 & $^{32}$S -- 94.99\%,	$^{33}$S -- 0.75\%, & \textgreater{}99\% \\
& &  &  &  $^{34}$S -- 4.25\%, $^{36}$S -- 0.01\% & \\
\bottomrule

\end{tabular}
\end{center}

\begin{multicols}{2}

The procedure and methods used to obtain and analyze the experimental data were discussed in detail in our previous article \cite{Fedorov2021}. Here we briefly outline their main content.

\begin{center}
\tabcaption{\label{table:adc} Characteristics of the digitizers used for data acquisition}
\footnotesize
\begin{tabular}{ l  l  l  l }
\toprule
Characteristic & ADCM-32 & CRS-6/16 \\
\hline
Number of channels & 32 & 6 \\
ADC bit depth & 14 bit & 11/16 bit \\
Sampling frequency & 66 MHz & 100 MHz \\
Input amplitude range & $-1 \div 1$ V & $-1 \div 1$ V \\
Data transfer rate & $\sim$250 MB/s & $\sim$190 MB/s \\
interface type & PCI-e & USB-3 \\
Max. count rate & $\sim10^5$ ev/s & $\sim5\times10^6$ ev/s \\
\bottomrule

\end{tabular}
\end{center}

\section{Data analysis}

The use of TNM made it possible to significantly reduce the contribution of the background events (correlated and random) to the recorded energy spectra of $\gamma$-quanta by organizing the coincidence of signals from $\gamma$-ray detectors and position-sensitive $\alpha$-detector. For the correct implementation of this technique, we carried out a thorough analysis of the time spectra of $\gamma-\alpha$ coincidences, which, following the terminology in neutron spectroscopy, can be called time-of-flight (TOF) spectra.

Figure~\ref{fig:TOF_Energy_BGO} shows different steps of analysis of the data obtained from the BGO-"Romasha" detector system. The upper figure shows the components of the TOF spectra: black curve (labeled A) -- total spectrum recorded with the sample; pink curve (labeled B) -- the background spectrum recorded in a separate experiment without sample, but with a sample holder and empty aluminum box; red curve (labeled C) -- difference between the two spectra. The background spectrum was normalized to the same total neutron flux as the spectrum with the sample. The first peak in all TOF spectra corresponds to the $\gamma$-rays while the second peak is due to the scattered neutrons. The components of the energy spectra (Fig.~\ref{fig:TOF_Energy_BGO}(b)) were constructed from the events which fall inside the corresponding windows in the TOF spectra: the $\gamma$-window was selected as 
$\pm2\sigma_{\gamma}$, where $\sigma_{\gamma}$ is the standard deviation of the Gaussian fit of the corresponding $\gamma$-peak, it is marked by pink vertical lines; the neutron window, marked by blue vertical lines, was chosen in such a way that events from $\gamma$-rays do not fall into it.

\begin{center}
\includegraphics[width=80mm]{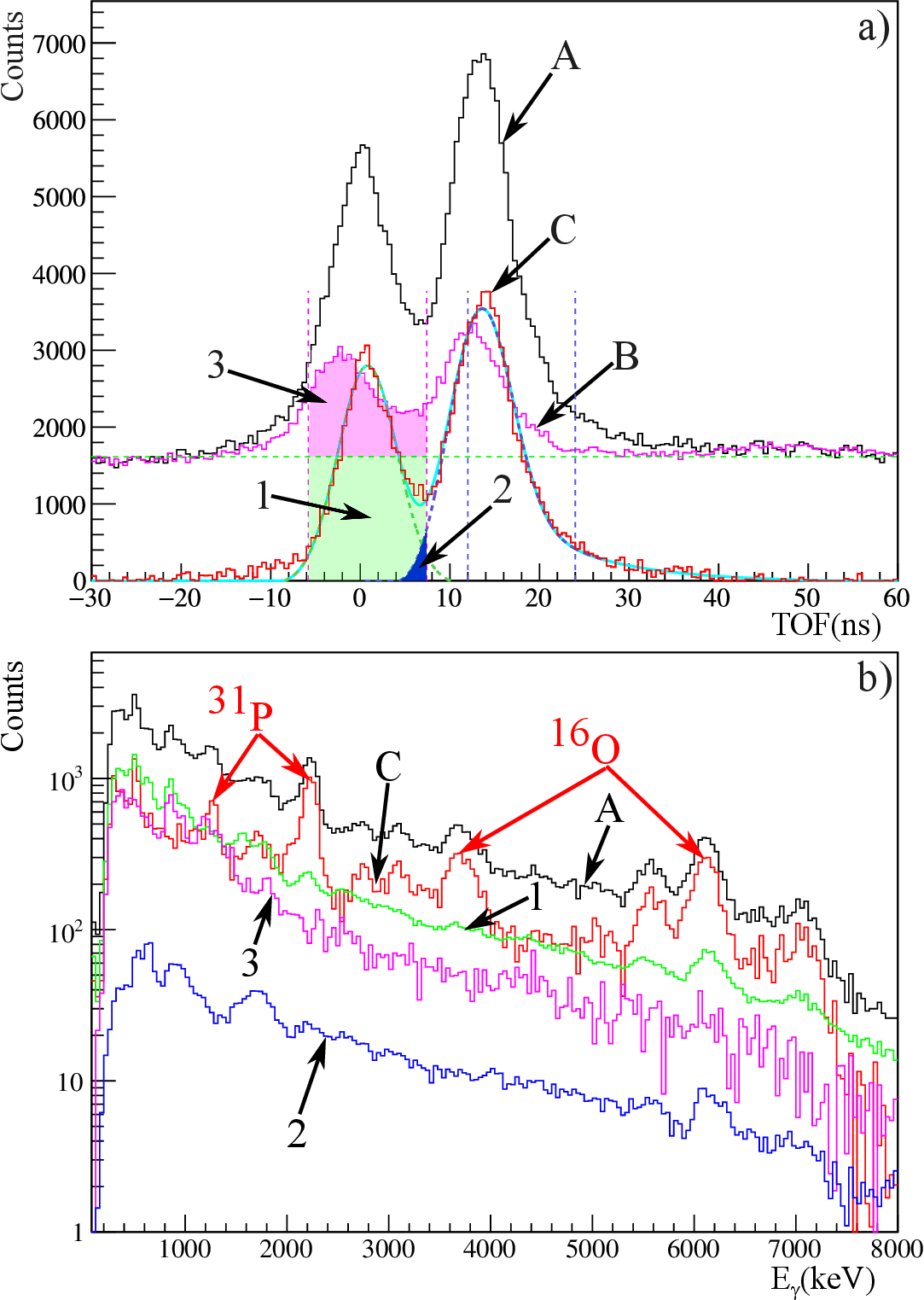}
\figcaption{\label{fig:TOF_Energy_BGO}\label{fig:TOF_Energy_BGO} An example of the TOF spectrum for the BGO detector $(a)$ and components of the energy spectra obtained at same detector $(b)$. See text for details.}
\end{center}

We identified three main sources of the background that must be taken into account when processing data: 
\begin{enumerate}
\item Random coincidences (labeled 1 in Fig.~\ref{fig:TOF_Energy_BGO}). It is constant in the time spectrum, and its height is determined by the counting rate in the detectors of $\alpha$-particles and $\gamma$-quanta involved in the coincidences. The background energy spectrum of the random coincidences was constructed by selecting events from the time spectrum to the left of the $\gamma$-peak. We used the time window from $-200$ to $-30$~ns. The random coincidence spectra were normalized by the ratio of the widths of the corresponding coincidence windows.

\item Neutrons scattered on the sample and its surrounding (labeled 2 in Fig.~\ref{fig:TOF_Energy_BGO}). Such events fall into the second peak in the TOF spectrum, but due to partial overlap with the $\gamma$-peak (blue region in Fig.~\ref{fig:TOF_Energy_BGO}(a)) these events also form a background in the $\gamma$-ray spectrum. The energy spectrum of the neutron background was determined in the neutron window, marked by blue vertical lines in Fig.~\ref{fig:TOF_Energy_BGO}(a) and normalized to the ratio of the corresponding components in the TOF spectra.

\item  Gamma-quanta formed during the interaction of neutrons by the environment near the sample (labeled 3 in Fig.~\ref{fig:TOF_Energy_BGO}).
The contribution of this background was determined from a separate measurement without sample. The energy spectrum of the $\gamma$-background was obtained in the same coincidence window of $\pm2\sigma_{\gamma}$ as the sample spectrum.
\end{enumerate}

The net energy spectrum, which is a difference between the total spectrum and the sum of all background components, was approximated by a special fitting function. Examples of the fit of the BGO energy spectra are shown in Fig.~\ref{fig:BGO_Spectr}. The fitting function includes a smooth background and a sum of response functions of the BGO detector for each $\gamma$-line, as described in \cite{Grozdanov2021}. The integral of the response function for the corresponding $\gamma$-line is proportional to the number of $\gamma$-quanta recorded by the detector and was used for determination of the angular distribution of the $\gamma$-ray emission.
\end{multicols}

\begin{center}
\includegraphics[width=160mm]{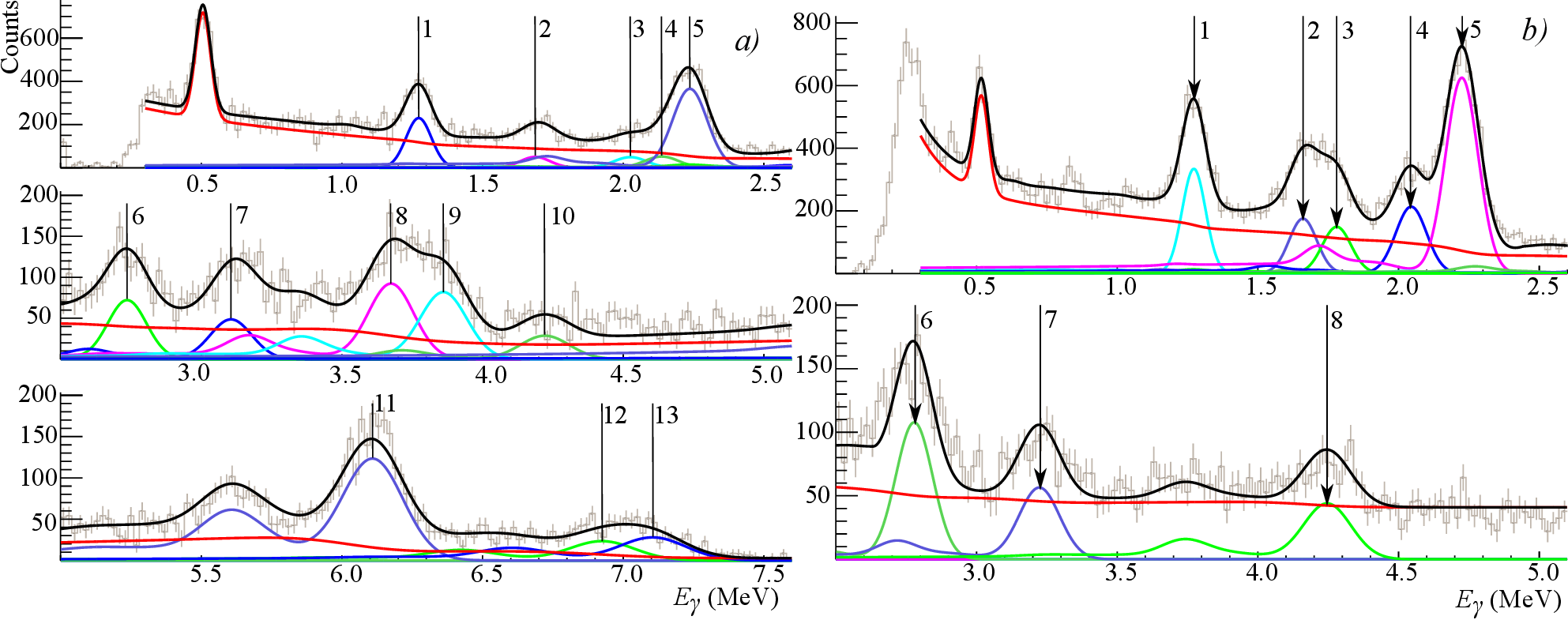}

\figcaption{\label{fig:BGO_Spectr} Energy spectra, obtained using a BGO-detector at an angle of $33^{\circ}$ with phosphorus oxide $(a)$ and sulfur $(b)$, approximated by the response function. 
Grey dots with error bars are the experimental data, solid black line is the full fitting function, red line is the sum of all background components, other colored lines are the response functions for the strongest characteristic $\gamma$-lines from oxygen, phosphorus and sulfur which are marked by vertical lines (see Tables \ref{table:Y_O}, \ref{table:Y_P} and \ref{table:Y_S}). }
\end{center}

\begin{multicols}{2}

The time resolution of the HPGe gamma detector does not allow separating useful events from background as good as by means of BGO scintillation detector.
At the same time, using the time-of-flight technique, it is still possible to isolate the time-constant background of random events generated mainly in the interaction of "untagged" neutrons with the environment. Due to the small distance from the sample and lower time resolution of the HPGe detector, the only peak in the time spectrum contains events from both $\gamma$-rays and secondary neutrons emitted from the target.

The time resolution of the HPGe detector shows strong dependence on the event energy \cite{Crespi2010}, therefore a special procedure was applied to select proper time window for each $\gamma$-ray energy and then improve background separation and prevent loss of good events.

We defined a 2-dimensional coincidence window, as shown in Fig.~\ref{fig:2dSpectrum} by approximating the 1-dimensional TOF profiles of the locus in different energy windows using the Gaussian fit. The width of the window in TOF-coordinate was defined as $\pm3\sigma$ of the Gaussian for each energy window. The random coincidence window which was used for the analysis of energy spectra is below the coincidence window having width of 300 ns.
    
According to our estimation, after the above described procedure the HPGe time resolution in different energy ranges is between $26$~ns (in the energy interval $600$ -- $900$~keV) and $17$~ns (in the interval $3600$ -- $3900$~keV). 

Figure~\ref{fig:P2O5_coin_anti} shows a fragment of energy distributions of $\gamma$-rays measured by the HPGe detector for a P$_2$O$_5$ sample for coincidence window (A), random coincidences (B) and the spectrum of "pure coincidences" (C), which is the difference between spectra A and B. Since the width of the coincidence window varies as a function of energy, and is different from the width of the random coincidence window, a special energy-dependent procedure was developed to normalize the random coincidence spectrum to the ratio of the corresponding window widths.

As can be seen, in the coincidence and "pure coincidence" spectra, the lines from the elements contained in the sample are significantly enhanced, and the background peaks are either absent or suppressed. 

After this procedure the number of registered events, which correspond to the $\gamma$-ray emission from different discrete states of nuclear reaction products, could be extracted from the $\gamma$-spectra. To obtain these values, the full energy absorption peaks were approximated by gaussians with linear background. 

\begin{center}
\includegraphics[width=82mm]{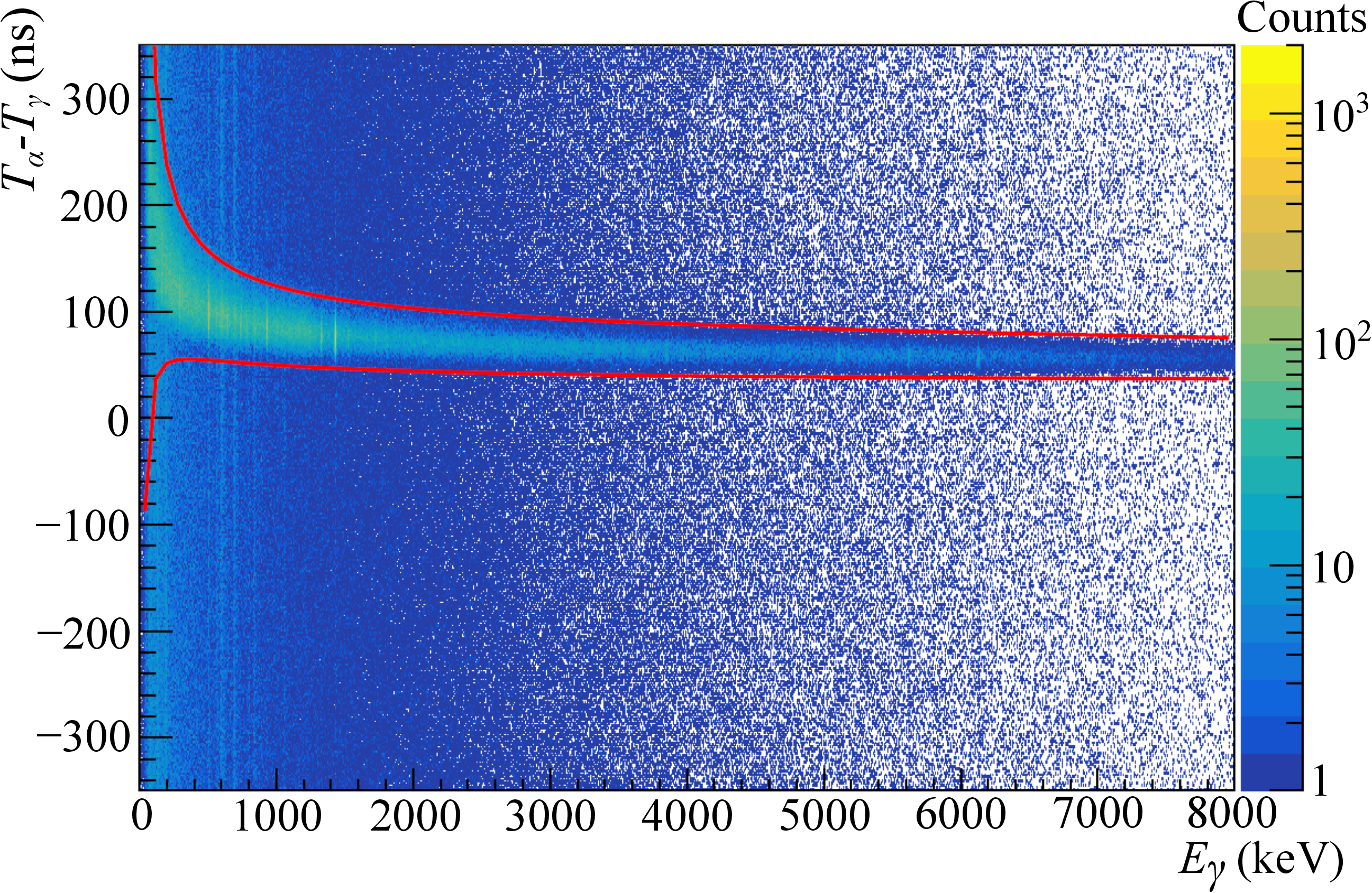}
\figcaption{\label{fig:2dSpectrum}Full TOF-Energy spectrum from measurement with HPGe detector. Color represents the number of counts. Coincidence window is marked by solid lines.}
\end{center}

In our experiments, we use relatively thick samples, which leads to tangible absorption and scattering of $\gamma$-rays and neutrons in the sample. Therefore, for each $\gamma$-detector a correction function was calculated by the Monte Carlo method, taking into account the re-scattering of neutrons in the sample, as well as the probabilities of absorption and detection of $\gamma$-quanta depending on their energies \cite{Grozdanov2021}.

\begin{center}
\includegraphics[width=82mm]{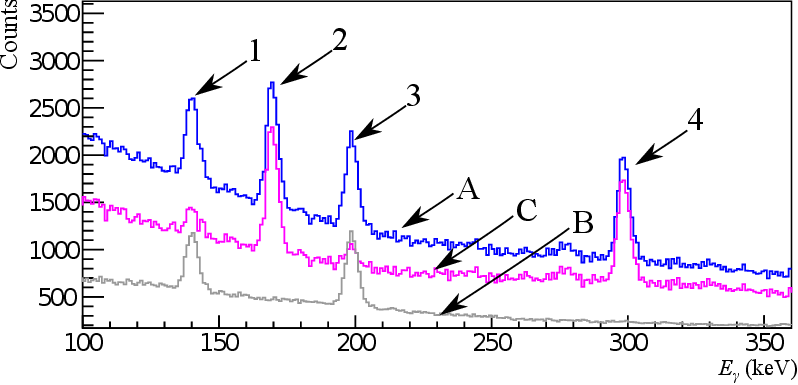}
\figcaption{\label{fig:P2O5_coin_anti} A fragment of the energy spectrum of $\gamma$-rays measured by the HPGe detector for a P$_2$O$_5$ sample. $A$ -- events in the coincidence window (blue line), $B$ -- background of random coincidences in the same window (grey line), $C$ -- difference (pure) spectrum ($A-B$) (magenta line). Characteristics of peaks indicated by numbers are: 1 -- $E_{\gamma}$ = $140$~keV from background, 2 -- $E_{\gamma}$ = $169.3$~keV from $^{16}$O$(n,\alpha$)$^{13}$C, 3 -- $E_{\gamma}$ = $200$~keV from background, 4 -- $E_{\gamma}$ = $298.2$~keV from $^{16}$O$(n,p)^{16}$N.}
\end{center}

\section{Results}
High-resolution $\gamma$-spectra have been obtained using the HPGe spectrometer. This allowed us to identify a large number of $\gamma$-lines and determine their yields.

Yields of $\gamma$-quanta $Y_{i}$, here and below are defined as the emission cross-sections for $\gamma$-quanta of a particular energy $E_{i}$, normalized to the cross section of the strongest (reference) transition ($E_{0}$) for a given element:

\begin{equation}
\label{eq:Yields_of_gamma}
Y_i = \frac{N_i}{N_0},
\end{equation}
where $N_i$ and $N_0$ are the numbers of emitted $\gamma$-quanta with an energy $E_i$ and $E_0$, respectively. The $\gamma$-quanta emission cross sections obtained in other experiments or calculated with the program TALYS 1.96 \cite{Koning2007} for comparison with our data were converted into yields using the following formula:

\begin{equation}
\label{eq:Yields_of_gamma_talys}
Y_i = \frac{\eta_j\sigma^{\gamma}_i}{\eta_0\sigma^{\gamma}_0},
\end{equation}
where $\eta_j$ -- the abundance of the $j$-th isotope, $\sigma^{\gamma}_i$ -- cross section for emission of $\gamma$-quanta with energy $E_i$, $\eta_0$ -- the abundance of the element with reference $\gamma$-transition, $\sigma^{\gamma}_0$ -- cross section for emission of $\gamma$-quanta with reference energy. 

Angular distributions of $\gamma$-quanta $W(\theta)$ obtained in our experiments using BGO detectors, were determined from the normalized number of $\gamma$-quanta in the detectors, corrected for absorption and scattering of $\gamma$-quanta and neutrons in the sample, according to the procedure described in \cite{Fedorov2021}. The coefficients of anisotropy in the angular distributions of $\gamma$-quanta are obtained as a result of the approximation of the angular distributions using the following expression:

\begin{equation}
\label{eq:legendre}
W(\theta)=1+\sum_{k=2,4...}^{2J}a_kP_k(\cos\theta).
\end{equation}

The obtained parameters of the angular anisotropy were used for correction of the yields of the strong $\gamma$-lines, for which the angular distribution was determined by us. The other $\gamma$-lines were assumed to be isotropic. 

\Correction{Systematic and statistical errors are taken into account in our analysis. The primary sources of systematic errors in our measurements are the uncertainties in the positions of the elements of the experimental setup resulting in changes of the geometrical efficiencies of the detectors and absorption of the $\gamma$-rays in the sample. The systematic errors were determined using Geant4 Monte Carlo simulations by varying the sample and detectors' positions in order to determine the differences in the corresponding experimental values. The resulting systematic errors are given in the corresponding tables.}

The TALYS code \cite{Koning2007} is widely used for theoretical calculations of various nuclear reactions induced by $n$, $p$, $d$, $^3$H, $^3$He, $\alpha$ in wide energy range. We have compared our results for $\gamma$-ray yields with values calculated by TALYS 1.96 using the "default" set of parameters.

The RIPL-3~\cite{RIPL-3} database used as a source of information about nuclear structure and properties for the nuclear reaction calculations in TALYS that makes it useful for the $\gamma$-spectra decoding. To simplify usage of the nuclear structure data from RIPL-3 and the TALYS calculation results in data processing programs we are developing a special software  named TalysLib. It is an object-orientied C++ library which uses ROOT~\cite{ROOT} capabilities for data drawing, storage and theoretical model parameters adjustment.

The main approach used in TALYS for neutron-induced reactions is optical model (OM) and its modifications implemented to describe inelastic processes: Distorted Wave Born Approximation (DWBA), symmetric rotational (ROT), harmonic vibrational (VIB), vibration-rotational model (VROT), asymmetric rotational (AROT) models (see \cite{Talys} for details). The choice of particular approach depends on the nuclear structure and available data on the nucleon scattering. For 54 nuclides there are predefined sets of optical model parameters in the TALYS nuclear structure. For other nuclei Koning parametrization \cite{Koning} is used.

For nuclei discussed in this paper predefined OM parameters and deformations available for $^{31}$P and $^{32}$S. For $^{16}$O they were calculated using Koning parameterization. Strictly speaking, light nuclei in the oxygen region lie at the limit of applicability of the Koning parameterization. However, as it was shown in \cite{Boromiza2020}, the calculations for oxygen are generally in good agreement with the experimental data.

The origin of the lowest excited states in $^{32}$S was investigated in several papers \cite{Mermaz1969,Haouat1984} and it was established that they have vibrational nature. By default for $^{32}$S in TALYS two bands with slightly different deformations are set. The first band consists of vibrational excited states with spin-parity $2^+_1$ and $0^+_2$, $2^+_2$, $4^+_1$, it has a deformation parameter $\beta = 0.29$. The state $3^-$ forms the second band with $\beta = 0.3$.

The OP parameterizations used in our calculations are given in the Appendix.

\subsection{Oxygen and Phosphorus}
The $\gamma$-ray yields of phosphorus and oxygen were obtained from measurements with phosphorus oxide P$_2$O$_5$. The parameters of the sample are given in Table~\ref{table:samples}. The high resolution spectrum of $\gamma$-quanta obtained in this experiment is shown in Fig. \ref{fig:P2O5_HPGe}, also showing the identified $\gamma$-lines; single (SE) and double (DE) escape peaks for the high-energy transitions are also marked. All lines were identified using the TalysLib library.

The yields of individual $\gamma$-transitions in the reactions of 14.1 MeV neutrons with oxygen nuclei were determined by formula~(\ref{eq:Yields_of_gamma}) and normalized to the yield of the most intense $\gamma$-line with an energy of 6129.9~keV. The results are presented in Table~\ref{table:Y_O}, which includes 8 transitions that we consider to be fairly reliably identified in our measurements, and 5 transitions cited in \cite{Simakov_1998} and present in the TALYS~1.9 database, but not observed in our spectra. For the lowest two $\gamma$-transitions ($E_{\gamma} = 169.3$ and 298.2~keV), the yields of $\gamma$-quanta were experimentally determined for the first time. For the $\gamma$-transitions shown in bold type, the angular distributions of $\gamma$-rays were also determined. The full errors (stat+sys) of $\gamma$-ray yields are indicated in brackets.

Gamma-quanta with an energy of 987~keV correspond to $\gamma$-transitions $1^-(7117\mbox{ keV})\rightarrow3^-(6130\mbox{ keV}))$, whereas branching ratio of this transition is significantly lower than that of the main transition $1^-(7117\mbox{ keV}))\xrightarrow{E1}0^+_{gs}$. Similarly, the transitions $E_\gamma$ = 1954.8~keV ($2^-(8872\mbox{ keV})\rightarrow2^+(6917\mbox{ keV})$) and 1755.1~keV ($2^-(8872\mbox{ keV})\rightarrow1^-(7117\mbox{ keV})$) have much lower probability than the transition with $E_\gamma$ = 2741.5~keV ($2^-(8872\mbox{ keV})\xrightarrow{E2}3^-(6130\mbox{ keV})$). The results of TALYS calculations generally confirm these ratios. The sensitivity of our experiment is not sufficient to measure characteristics of transitions with such a low probability.

We should note the discrepancy in the results of TALYS model calculations for $E_\gamma=6917.1$~keV ($2^+\rightarrow0^+_{g.s.}$). The overestimated values of the model calculations were pointed out by the authors \cite{Boromiza2020}. As for the discrepancies in the description of the ($n,\alpha$) channels, they are not so significant and can be attributed to the difficulties of the model description of this type of reactions in TALYS. 

\end{multicols}

\begin{center}
\includegraphics[width=160mm]{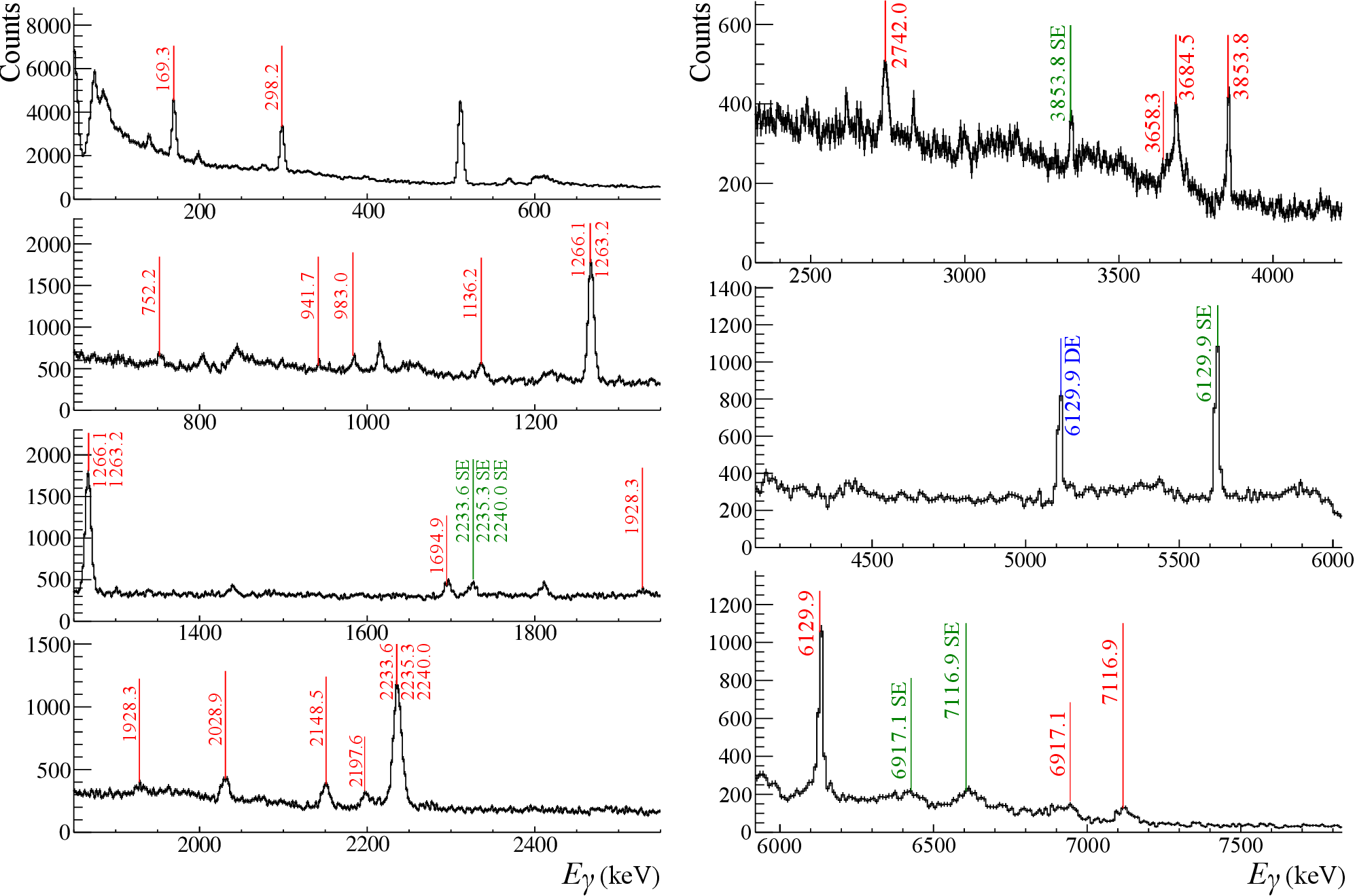}
\figcaption{\label{fig:P2O5_HPGe} Energy spectrum of $\gamma$-quanta from the P$_2$O$_5$ sample in the range $50-7600$~keV, measured by the HPGe detector. Position of $\gamma$-peaks identified by TalysLib and attributed to reactions with oxygen or phosphorus are marked by red lines. Single escape (SE) and double escape (DE) peaks of strong $\gamma$-transitions are marked by green and blue lines, respectively. \Correction{The peak at 511 keV corresponds to electron-positron annihilation processes.}}
\end{center}

\begin{multicols}{2}
The yields of $\gamma$-rays  for phosphorus were normalized to the sum of three lines $2233.6$, $2235.3$ and $2240.0$~keV, which couldn't be separated in our experiment. The results are summarized in Table~\ref{table:Y_P}. For 9 $\gamma$-transitions the yields of individual $\gamma$-lines were determined experimentally for the first time. The most intense  $\gamma$-transitions for which the angular distributions are measured are marked in bold.

Overall, the agreement in yields between our results and literature, as well as with the calculated data can be considered satisfactory. However, the spectroscopy of odd-even nuclei has somewhat more complicated structure. In the case of $^{31}$P, some peaks are overlapped with $\gamma$-lines from other reactions, such as $(n,p)$, $(n,d)$ and $(n,\alpha)$. In particular, the strongest $\gamma$-line, which is identified in \cite{Simakov_1998} as the second excited state to ground state $(\frac{5}{2}^+ (2234\mbox{ keV})\rightarrow {\frac{1}{2}}^+(g.s.))$ transition in $^{31}$P from the $(n,n')$ reaction, according to our calculations by TALYS, is a triplet with the dominant $\gamma$-line 2235.3~keV, which is the first excited state to ground state $(2^+ \rightarrow 0^+_{gs}$ transition, emitted from $^{30}$Si after the $^{31}$P$(n,d)^{30}$Si reaction. At the same time, the second strongest peak at 1266.1~keV, which is the first excited state to ground state $(\frac{3}{2}^+ \rightarrow {\frac{1}{2}}^+_{gs}$ transition in $^{31}$P has an admixture of the $(2^+ (3499\mbox{ keV})\rightarrow 2^+ (2235\mbox{ keV})$) transition in $^{30}$Si.
This circumstance significantly complicates the choice of the transition for normalization when determining reaction yields.

\end{multicols}

\begin{center}
\tabcaption{\label{table:Y_O} 
Experimental $\gamma$-ray yields from oxygen in comparison with 
with TALYS 1.96 calculations and published data of other authors. \Correction{Statistical uncertainties are indicated in brackets. Estimated systematic uncertainties do not exceed 5\% of the yield values.} For $\gamma$-transitions in bold type, the angular distributions of $\gamma$-quanta were also determined. The numbers in the first column indicate the peaks in the energy spectra obtained with the Romasha BGO detector system (see Fig.~\ref{fig:BGO_Spectr} (a)).}

\footnotesize
\begin{tabular}{l l  l  l l  l  l }

\toprule 
No &{$E_{\gamma}$, keV} & 
{Reaction} & 
\multicolumn{4}{c}{$Y_{\gamma}$, \%} \\ \cline{4-7}
&
 &
 &
{This work} &
{TALYS} &
{\cite{Simakov_1998}} &
{\cite{Nyberg-Ponnert}} \\
\hline

&169.3 & $^{16}$O$(n,\alpha)^{13}$C & 32.9 (3.7) & 16.0 & & \\
\hline
&298.2 & $^{16}$O$(n,p)^{16}$N & 43.4 (3.9) & 15.6 & & \\
\hline
&987.0 & $^{16}$O$(n,n')^{16}$O & - & - & 4.2 (0.8) & \\
\hline
&1755.1 & $^{16}$O$(n,n')^{16}$O & - & 5.2 & 4.6 (0.9) & \\
\hline
&1954.8 & $^{16}$O$(n,n')^{16}$O & - & 1.6 & 4.1 (2.6) & \\
\hline
6 & \textbf{2742.0} & \textbf{$^{16}$O$(n,n')^{16}$O} & \textbf{34.7 (4.3)} & \textbf{35.1} & \textbf{25.7 (3.2)} & \textbf{31.3 (7.0)} \\
\hline
7 & 3089.4 & $^{16}$O$(n,\alpha)^{13}$C & - & 20.7 & 14.9 (1.7) & \\
\hline
8 & \textbf{3684.5} & \textbf{$^{16}$O$(n,\alpha)^{13}$C} & \textbf{35.9 (6.3)} & \textbf{50.5} & \textbf{38.9 (4.5)} & \textbf{29.6 (6.2)} \\
\hline
9 & \textbf{3853.8} & \textbf{$^{16}$O$(n,\alpha)^{13}$C} & \textbf{33.0 (4.1)} & \textbf{27.6} & \textbf{22.8 (3.3)} & \textbf{23.5 (4.9)} \\
\hline
10 & 4438.9 & $^{16}$O$(n,n'\alpha)^{12}$C & - & - & 11.6 (1.7) & \\
\hline
11 & \textbf{6129.9} & \textbf{$^{16}$O$(n,n')^{16}$O} & \textbf{100} & \textbf{100} & \textbf{100} & \textbf{100} \\
\hline
12 & 6917.1 & $^{16}$O$(n,n')^{16}$O & 31.9 (3.9) & 82.9 & 31.8 (3.7) & 34.8 (7.3) \\
\hline
13 & 7116.9 & $^{16}$O$(n,n')^{16}$O & 34.5 (5.1) & 32.6 & 36.1 (4.4) & 39.1 (8.4) \\
\bottomrule

\end{tabular}
\end{center}

\begin{center}
\tabcaption{\label{table:Y_P} Experimental $\gamma$-ray yields from phosphorus, see Table~\ref{table:Y_O} for explanations.}

\footnotesize
\begin{tabular}{ l l  l  l  l  l }
\toprule
No & \multicolumn{1}{c}{$E_{\gamma}$, keV} & 
\multicolumn{1}{c}{Reaction} & 
\multicolumn{3}{c}{$Y_{\gamma}$, \%} \\ \cline{4-6}

 & \multicolumn{1}{c}{} &
\multicolumn{1}{c}{} &
\multicolumn{1}{c}{This work} &
\multicolumn{1}{c}{TALYS} &
\multicolumn{1}{c}{\cite{Simakov_1998}} \\
\hline

 & 752.2 & $^{31}$P$(n,p)^{31}$Si & 5.1 (2.1) & 3.6 & \\
\hline
 & 983.0 & $^{31}$P$(n,\alpha)^{28}$Al & 4.1 (0.9) & 5.0 & \\
\hline
 & 1136.2 & $^{31}$P$(n,n')^{31}$P & 6.9 (1.2) & 1.7 & \\
\hline
 \multirow{2}{*}{1} & \textbf{1263.3$^*$} & \textbf{$^{31}$P$(n,d)^{30}$Si} &  \multirow {2 }{*}{\textbf{{56.9 (9.0)}}} & \textbf{3.9} & \\
 & \textbf{1266.1$^*$} & \textbf{$^{31}$P$(n,n')^{31}$P} &  & \textbf{42.0} & \textbf{43.8 (10.6)} \\
\hline
  & 1438.6 & $^{31}$P$(n,p)^{31}$Si & 5.5 (1.8) & 3.2 & \\
\hline
2 & 1694.9 & $^{31}$P$(n,p)^{31}$Si & 11.2 (2.2) & 6.7 & \\
\hline
  & 1928.3 & $^{31}$P$(n,n')^{31}$P & 5.5 (1.8) & 3.0 & \\
\hline
3 & 2028.9 & $^{31}$P$(n,n')^{31}$P & 14.8 (2.5) & 12.7 & \\
\hline
4 & 2148.5 & $^{31}$P$(n,n')^{31}$P & 14.0 (3.0) & 12.7 & 14.6 (3.3)\\
\hline
 & 2197.6 & $^{31}$P$(n,n')^{31}$P & 5.4 (1.1) & 2.3 & \\
\hline
\multirow{3}{*}{5} & \textbf{2233.6$^*$} & \textbf{$^{31}$P$(n,n')^{31}$P} & \multirow{3}{*}{\textbf{{100}}} & \textbf{18.9} &  \multirow{3}{*}{\textbf{{100}}}\\
& \textbf{2235.3$^*$} & \textbf{$^{31}$P$(n,d)^{30}$Si} &  & \textbf{78.1} &  \\
 & \textbf{2240.0$^*$} & \textbf{$^{31}$P$(n,n')^{31}$P} & & \textbf{2.9} & \\
\hline
& 3658.3 & $^{31}$P$(n,n')^{31}$P & 12.6 (2.9) & 1.3 & \\
\hline

\end{tabular}
\end{center}

\begin{multicols}{2}
The angular distributions $W(\theta)$ determined for the strongest $\gamma$-transitions corresponding to reactions of fast neutrons with oxygen are shown in Fig.~\ref{fig:W_O}, and with phosphorus in Fig.~\ref{fig:W_P}. 
The parameters of the Legendre polynomials used to approximate $W(\theta)$ are given in Table~\ref{table:W_O} for oxygen and in Table~\ref{table:W_P} for phosphorus in comparison with the results of other published works, as well as with ENDF/B-VIII.0 data \cite{ENDF}. The anisotropy parameters for the literature data, if not given by the authors, were calculated by us from the published angular distributions using Eq.~(\ref{eq:legendre}).

\end{multicols}

\begin{center}
\includegraphics[width=150mm]{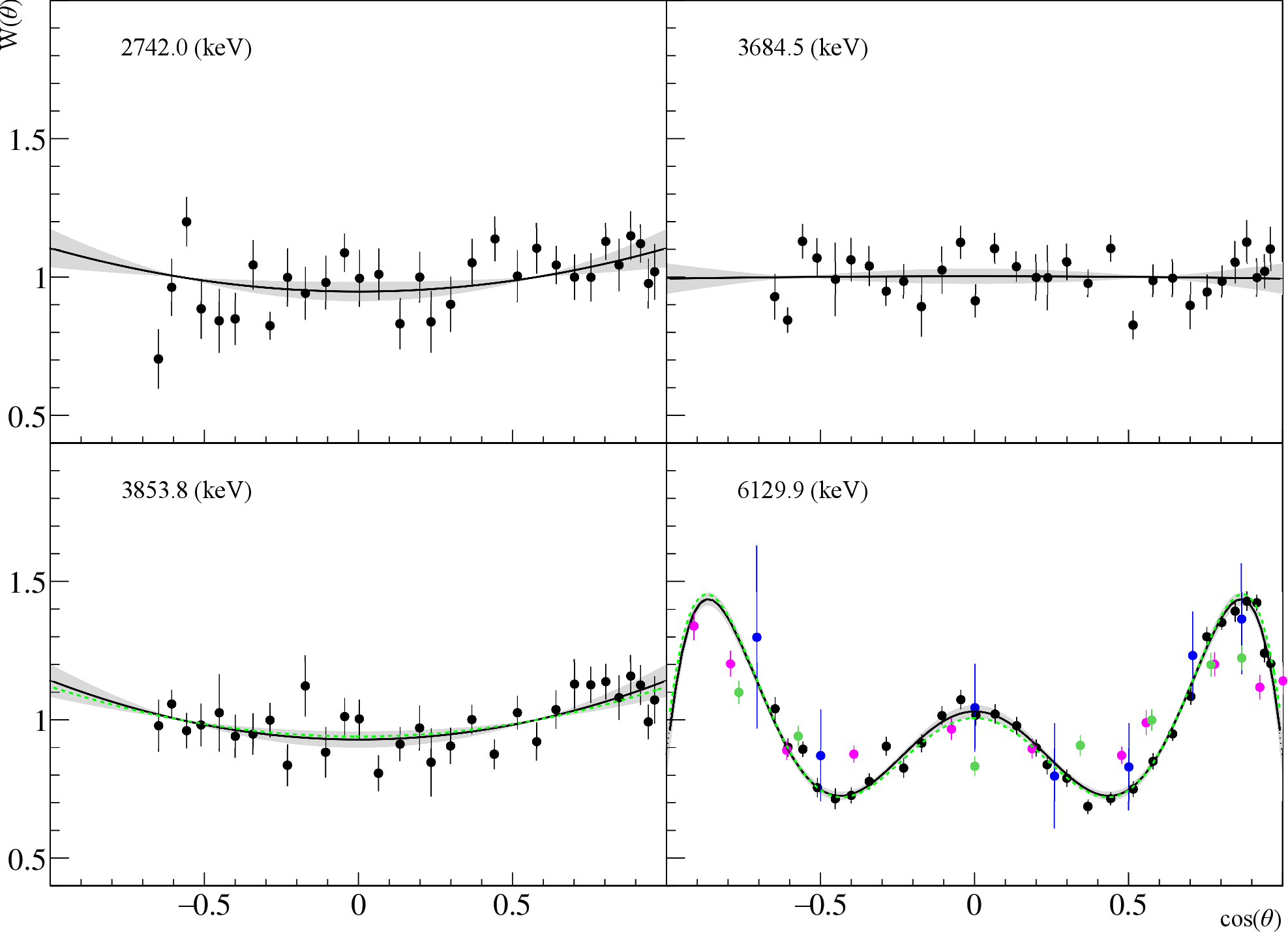}
\figcaption{\label{fig:W_O} Angular distributions of $\gamma-$quanta for transitions in oxygen. Black dots correspond to data from this work, blue -- from \cite{Kozlowski1965}, purple -- \cite{Morgan1964}, green -- \cite{McDonald1966}. Black solid line is the Legendre polynomial fit of data from this work using formula (\ref{eq:legendre}); green dashed line is the angular distribution from ENDF/B-VIII  \cite{ENDF}. The grey area indicates the 95\% confidence interval for the approximation of our data, taking into account their statistical and systematic errors of angular anisotropy parameters.}
\end{center}

\begin{center}
\tabcaption{\label{table:W_O} 
Expansion coefficients of Legendre polynomials for angular distributions of $\gamma$-quanta from radiative transitions in oxygen. \Correction{Statistical and systematical uncertainties are indicated in brackets at top and bottom, respectively.} The energies of $\gamma$-quanta $E_{\gamma}$, as well as the spin, parity and energy $J^P(E)$ of the initial ($i$) and final ($f$) states are taken from \citep{Tilley1993}.}
\footnotesize
\begin{tabular}{ l  l  l  l  l  l  l  l }

\toprule 
{$E_{\gamma}$, keV} & 
{Reaction} & 
{{$J_i^P$  }} &
{{$J_f^P$  }} &
{Ref.} &
{$a_2$} &
{$a_4$} &
{$a_6$} \\
 &
 &
($E_i$, keV) &
($E_f$, keV) &
 &
 &
 &
 \\

\hline

2742.0 & $^{16}$O$(n,n')^{16}$O & 2$^-$ (8871.9) & 3$^-$ (6129.9) & This work & 0.10 \Correction{$\left(0.04\atop0.01\right)$} & * &  \\
\hline
3684.5 & $^{16}$O$(n,\alpha)^{13}$C & $\dfrac{3}{2}^-$ (3684.5) & $\dfrac{1}{2}^-$ (0) & This work & $-0.01$ \Correction{$\left(0.04\atop0.01\right)$}  & &  \\

\hline

\multirow{2}{*}{3853.8} & \multirow{2}{*}{$^{16}$O$(n,\alpha)^{13}$C} & \multirow{2}{*}{$\dfrac{5}{2}^+$ (3853.8)} & \multirow{2}{*}{$\dfrac{1}{2}^-$ (0)} & \cite{ENDF} & 0.12 & & \\
& & & & This work  &  0.14 \Correction{$\left(0.03\atop0.01\right)$} & &  \\
\hline

\multirow{6}{*}{6129.9} & \multirow{6}{*}{$^{16}$O$(n,n')^{16}$O} & \multirow{6}{*}{3$^-$ (6129.9)} & \multirow{6}{*}{0$^+$ (0)} & \cite{ENDF} & 0.40 & 0.09 & $-0.55$ \\
& & & & \cite{Morgan1964} & 0.35 (0.07) & 0.02 (0.08) & $-0.03$ (0.08) \\
& & & & \cite{Kozlowski1965} & 0.18 (0.06) & $-0.27$ (0.08) & $-0.68$ (0.08) \\
& & & & \cite{McDonald1966} & 0.26 (0.04) & 0.06 (0.05) & $-0.21$ (0.05)  \\
& & & & \cite{Grozdanov2018} & 0.34 (0.02) & 0.10 (0.02) & $-0.26$ (0.08)  \\
& & & & This work & 0.36 \Correction{$\left(0.02\atop0.01\right)$} & 0.08 \Correction{$\left(0.02\atop0.03\right)$} & $-0.57$ \Correction{$\left(0.03\atop0.03\right)$}\\
\bottomrule

\end{tabular}
\end{center}

\begin{multicols}{2}
It should be noted that experimental literature data for the $\gamma$-ray angular anisotropy in oxygen exist only for the strongest $6129.9$~keV line; for the remaining three transitions, the angular correlation coefficients were obtained for the first time. Overall, we can note acceptable agreement of our data both with literature and with library data. 

The anisotropy parameters obtained in this experiment for the 6129.9~keV transition in oxygen are comparable with the results of our previous work where measurements were performed with a SiO$_2$ target \cite{Grozdanov2018}.

For the transitions 2742.0 keV and 3684.5 keV in oxygen the isotropic distribution of $\gamma$-quanta is indicated in the ENDF/B-VIII library, which agrees well with our data (falling within the 95\% confidence interval of our approximation). 

In the case of phosphorus the strongest $\gamma$-lines 1266.1 keV and 2235.3 keV couldn't be separated from the neighboring transitions, therefore the angular anisotropy parameters in Table~\ref{table:W_P} are given for the sum of unresolved transitions. We have not found published experimental data on the angular anisotropy of $\gamma$-ray emission in neutron-induced reactions on phosphorus. ENDF/B-VIII library also lacks data on the evaluated $\gamma$-ray angular anisotropy. Therefore, we cannot compare our results presented in Table~\ref{table:W_P} with any other data.

\end{multicols}

\begin{center}
\includegraphics[width=150mm]{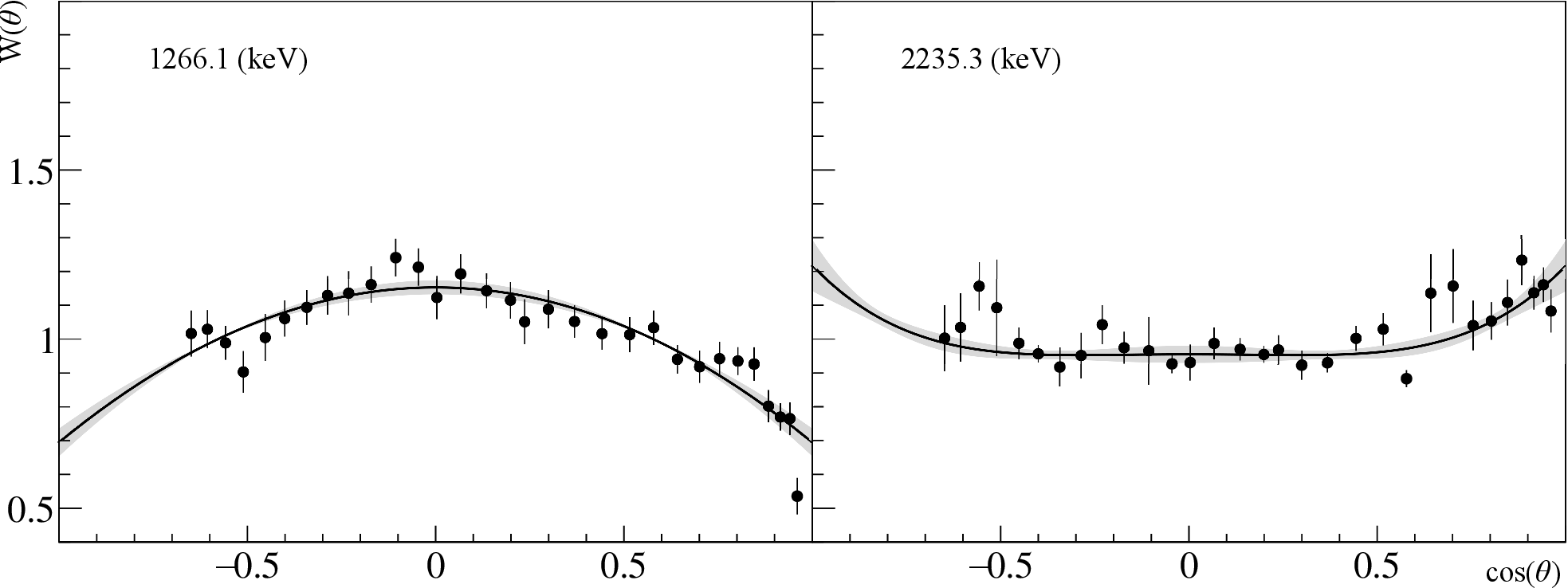}
\figcaption{\label{fig:W_P}
Angular distributions of $\gamma$-quanta from radiative transitions in phosphorus. The energies correspond to the strongest transition in the unresolved peaks. Black dots show the data of this work. The black solid line is the approximation of these data by the Legendre polynomials according to formula (\ref{eq:legendre}).
The grey areas show the 95\% confidence interval of the approximation made, taking into account the statistical and systematic errors of the angular anisotropy parameters.}
\end{center}

\begin{center}
\tabcaption{\label{table:W_P} 
Expansion coefficients of Legendre polynomials for angular distributions of $\gamma$-quanta from radiative transitions in phosphorus. \Correction{Statistical and systematical uncertainties are indicated in brackets at top and bottom, respectively.} The energies of $\gamma$-quanta $E_{\gamma}$, as well as the spin, parity and energy $J^P(E)$ of the initial ($i$) and final ($f$) states are taken from \citep{Tilley1993}.}

\footnotesize
\begin{tabular}{c  c  c  c  c  c  c  c }

\toprule 
{$E_{\gamma}$, keV} & 
{Reaction} & 
{$J_i^P$} &
{$J_f^P$} &

{Ref.} &
{$a_2$} &
{$a_4$} \\
 &
&
($E_i$, keV) &
($E_f$, keV) &
&  & \\

\hline
1263.3& $^{31}$P$(n,d)^{30}$Si & 2$^+$ (3498.5) & 2$^+$ (2235.3) & \multirow{2}{*}{$
\begin{cases}
\mbox{This work}
\end{cases}$} & \multirow{2}{*}{$-0.31$ \Correction{$\left(0.02\atop0.01\right)$}} &  \\
1266.1& $^{31}$P$(n,n')^{31}$P & $\dfrac{3}{2}^+$ (1266.1) & $\dfrac{1}{2}^+$ (0) &  &  &  \\
\hline
2233.6 & $^{31}$P$(n,n')^{31}$P & $\dfrac{5}{2}^+$ (2233.6) & $\dfrac{1}{2}^+$ (0) & & & \\
2235.3 & $^{31}$P$(n,d)^{30}$Si & 2$^+$ (2235.3) & 0$^+$ (0) & $\begin{cases} \\ \mbox{This work} \\ \\ \end{cases} $ & 0.14 \Correction{$\left(0.02\atop0.01\right)$}  & 0.07 \Correction{$\left(0.03\atop0.04\right)$}\\
2240.0 & $^{31}$P$(n,n')^{31}$P & $\dfrac{3}{2}^+$ (3506.1) & $\dfrac{3}{2}^+$ (1266.1) &  & & \\
\hline

\end{tabular}
\end{center}

\newpage

\begin{multicols}{2}

\subsection{Sulfur}
The high resolution energy spectrum of $\gamma$-quanta from sulfur obtained with a HPGe detector is shown in Fig.~\ref{fig:S}. Vertical lines point to the observed $\gamma$-transitions, attributed to the reactions of 14.1 MeV neutrons with sulfur and identified by TalysLib.

The $\gamma$-ray yields of sulfur were determined by formula~(\ref{eq:Yields_of_gamma}) and normalized to the sum of three lines: $2228.5$, $2230.6$ and $2233.6$~keV. Although the transition to the ground state $2^+ (2230.6\mbox{ keV})\rightarrow0^+_{gs}$ is dominant, the resolution of our detector was not sufficient to separate it from the neighboring $\gamma$-lines. Table~\ref{table:Y_S} includes 10 single $\gamma$-transitions, two unresolved doublets and one unresolved triplet that were identified in our experiment. The experimental yields  we obtained are compared to the TALYS calculations and literature data \cite{Simakov_1998}.

In our work for 8 individual $\gamma$-transitions and one doublet the $\gamma$-ray yields were determined experimentally for the first time. For $\gamma$-transitions in bold, the angular distributions were also determined. In general, there is a reasonable agreement between our results and literature data for strong $\gamma$-transitions, except for the $\gamma$-line $E_{\gamma} = 2028.2\mbox{ keV}$. We believe that the reason for this discrepancy is the admixture of lines 2127.6~keV and 2217.8~keV in the data from \cite{Simakov_1998}, which were obtained by a low resolution NaI(Tl) $\gamma$-detector. The comparison of our results with TALYS calculations is not perfect, but the accuracy of TALYS data can be quite poor, especially for the reactions $(n,p)$, $(n,d)$ and $(n,\alpha)$.

The angular distributions $W(\theta)$, determined for the strongest gamma transitions from reactions induced by neutrons with an energy of 14.1 MeV on a sulfur sample, are shown in Fig.~\ref{fig:W_S}. The obtained parameters of the angular distributions are given in Table~\ref{table:W_S} in comparison with the results of other works.

It can be seen that the errors in our data presented in Fig.~\ref{fig:W_S} are smaller than in the data of other authors. It also resulted in smaller errors in the angular distribution coefficients shown in Table~\ref{table:W_S}. Data from the evaluated data library are not presented in the table because angular distributions for these transitions in ENDF/B-VIII are treated as isotropic.

\begin{center}
\includegraphics[width=70mm]{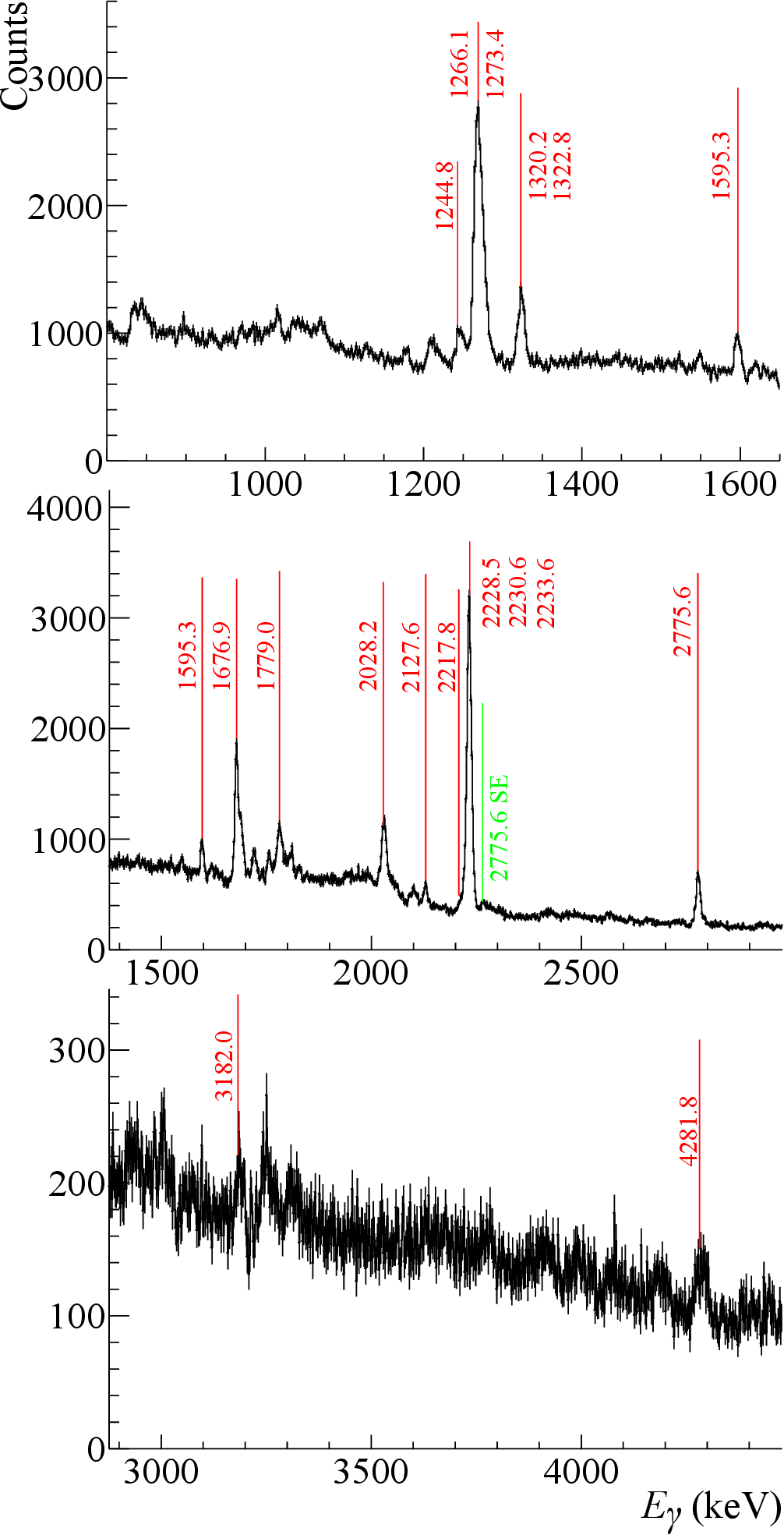}
\figcaption{\label{fig:S}  Energy spectrum of $\gamma$-quanta in the range $800-4500$~keV, measured with the HPGe detector during the irradiation of a sulfur sample. Green line -- single escape peak. Identification of the $\gamma$-peaks was performed via TalysLib, see details in text.}
\end{center}

\end{multicols}

\newpage

\begin{center}
\tabcaption{\label{table:Y_S} Comparison of experimental yields of $\gamma$-rays from sulfur obtained by the HPGe detector with TALYS 1.96 calculations and published data of other authors. \Correction{Statistical uncertainties are indicated in brackets. Estimated systematic uncertainties do not exceed 5\% of the yield values.} For $\gamma$-transitions in bold type, the angular distributions of $\gamma$-quanta were also determined. An asterisk (*) marks unresolved $\gamma$- transitions. The numbers in the first column indicate these peaks in the energy spectra obtained with the Romasha BGO detector system (see Fig.~\ref{fig:BGO_Spectr}(b)). }

\footnotesize
\begin{tabular}{l l  l  l  l  l }
 
\toprule 
No & \multicolumn{1}{c}{{$E_{\gamma}$, keV}} & 
\multicolumn{1}{c}{Reaction} & 
\multicolumn{3}{c}{$Y_{\gamma}$, \%} \\ \cline{4-6}

 & \multicolumn{1}{c}{} &
\multicolumn{1}{c}{} &
\multicolumn{1}{c}{This work} &
\multicolumn{1}{c}{TALYS} &
\multicolumn{1}{c}{\cite{Simakov_1998}} \\
\hline

 & 1244.8 & $^{32}$S$(n,p)^{32}$P & 5.3 (1.0) & 3.8 & \\
\hline
\multirow{2}{*}{1} & \textbf{1266.1$^*$} & \textbf{$^{32}$S$(n,d)^{31}$P} & \multirow{2}{*}{\textbf{{40.1 (1.6)}}} & \textbf{35.4} &  \\
 & \textbf{1273.4$^*$} & \textbf{$^{32}$S$(n,\alpha)^{29}$Si} & & \textbf{31.6} & \textbf{43.6 (7.0)}\\
\hline
 & 1320.2$^*$ & $^{34}$S$(n,n')^{34}$S & \multirow{2}{*}{{10.2 (3.7)}} & 1.5 & \\
 & 1322.8$^*$ & $^{32}$S$(n,p)^{32}$P & & 5.5 & \\
\hline
 & 1595.3 & $^{32}$S$(n,\alpha)^{29}$Si & 6.4 (0.5) & 11.0 & \\
\hline
2 & 1676.9 & \Correction{$^{32}$S$(n,p)^{32}$P} & 24.7 (1.0) & 18.7 & \\
\hline
3 & 1779.0 & $^{32}$S$(n,n\alpha)^{28}$Si & 9.3 (0.9) & 16.8 & \\
\hline
4 & \textbf{2028.2} & \textbf{$^{32}$S$(n,\alpha)^{29}$Si} & \textbf{18.1 (1.3)} & \textbf{40.1} & \textbf{39.5 (9.6)}\\
\hline
 & 2127.6 & $^{34}$S$(n,n')^{34}$S & 4.7 (0.4) & 9.1 & \\
\hline
 & 2217.8 & $^{32}$S$(n,p)^{32}$P & 7.1 (0.6) & 1.7 & \\
\hline
\multirow{3}{*}{5}& \textbf{2228.5$^*$} & \textbf{$^{32}$S$(n,n')^{32}$S} & \multirow{3}{*}{\textbf{{100}}} & \textbf{10.4} & \\
 & \textbf{2230.6$^*$} & \textbf{$^{32}$S$(n,n')^{32}$S} &  & \textbf{79.0} & \textbf{{100}}\\
 & \textbf{2233.6$^*$} & \textbf{$^{32}$S$(n,d)^{31}$P} & & \textbf{10.6} & \\
\hline
6 & \textbf{2775.6} & \textbf{$^{32}$S$(n,n')^{32}$S} & \textbf{17.0 (1.1)} & \textbf{18.5} & \textbf{14.8 (2.4)} \\
\hline
7 & 3182.0 & $^{32}$S$(n,n')^{32}$S & 4.1 (1.0) & 3.0 & \\
\hline
8 & 4281.8 & $^{32}$S$(n,n')^{32}$S & 5.4 (1.0) & 6.1 & \\
\bottomrule

\end{tabular}
\end{center}

\begin{center}
\includegraphics[width=150mm]{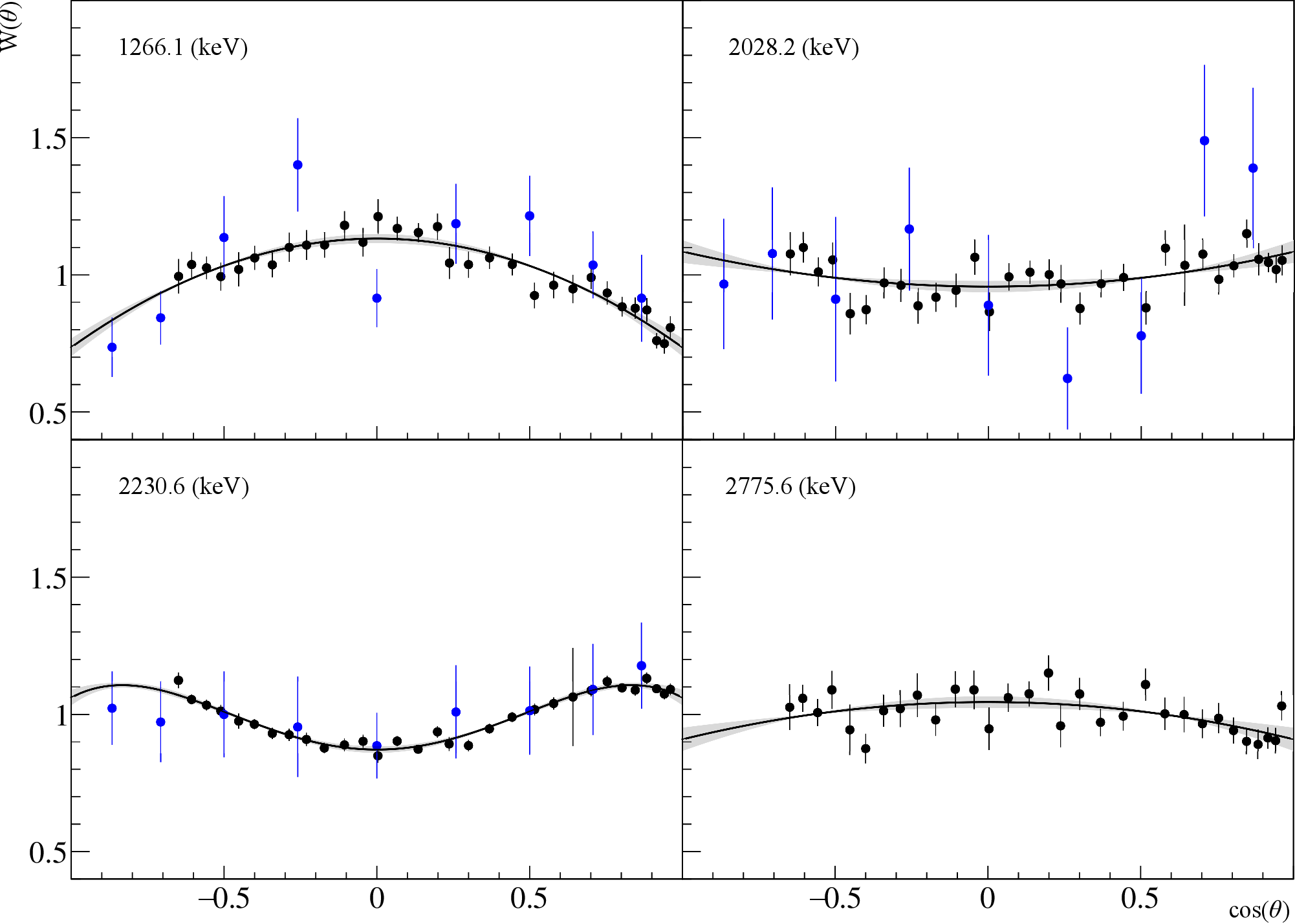}
\figcaption{\label{fig:W_S} Angular distributions of $\gamma$-quanta from radiative transitions in sulfur. The energies correspond to the strongest transition in the unresolved peaks. Black dots show the data of this work, blue -- data from \cite{Abbondanno1973}. The black solid line is the polynomial approximation of our data by the Legendre polynomials according to Eq.~(\ref{eq:legendre}).
The gray areas show the 95\% confidence interval of the approximation made, taking into account the statistical and systematic errors of the angular anisotropy parameters.}
\end{center}

\begin{center}
\tabcaption{\label{table:W_S} Legendre polynomial expansion coefficients for the angular distribution of $\gamma$-quanta for sulfur. \Correction{Statistical and systematical uncertainties are indicated in brackets at top and bottom, respectively.} The $\gamma$-quanta energies $E_{\gamma}$ and also the spin, parity and energy $J^P$(E) of the initial ($i$) and final ($f$) states are taken from \citep{Tilley1993}.}
\footnotesize
\begin{tabular}{ c  c  c  c  c  c  c c }
 
\toprule 
{$E_{\gamma}$, keV} & 
{Reaction} & 
{$J_i^P$} &
{$J_f^P$} &

{Ref.} &
{$a_2$} &
{$a_4$} \\
 &
 &
($E_i$, keV) &
($E_f$, keV) &
 &
 &
 \\

\hline

1266.1 & $^{32}$S$(n,d)^{31}$P & $\dfrac{3}{2}^+$ (1266.1) & \multirow{2}{*}{$\dfrac{1}{2}^+$ (0)} & \multirow{2}{*}{$\begin{cases} \text{\cite{Abbondanno1973}}\\ \text{This work}\end{cases}$ } & $-0.28$ (0.13) & \\

1273.4& $^{32}$S$(n,\alpha)^{29}$Si & $\dfrac{3}{2}^+$ (1273.4) & &   &$-0.26$ \Correction{$\left(0.02\atop0.01\right)$}  & \\

\hline
\multirow{2}{*}{2028.2} & \multirow{2}{*}{$^{32}$S$(n,\alpha)^{29}$Si} & \multirow{2}{*}{$\dfrac{5}{2}^+$ (2028.2)} & \multirow{2}{*}{$\dfrac{1}{2}^+$ (0)}  & \multirow{2}{*}{$\begin{cases}\text{\cite{Abbondanno1973}}\\ \mbox{This work}\end{cases}$ } &  0.35 (0.20)& \\
 & & & &  & 0.08 \Correction{$\left(0.02\atop0.01\right)$} &\\
\hline
2228.5 & $^{32}$S$(n,n')^{32}$S & 4$^+$ (4459.1) & 2$^+$ (2230.6) & \multirow{3}{*}{$\begin{cases}\text{\cite{Abbondanno1973}}\\ \mbox{This work} \\ \\ \end{cases}$ } & 0.13 (0.05)  &$-0.04$ (0.08)  \\
2230.6 & $^{32}$S$(n,n')^{32}$S & 2$^+$ (2230.6) & 0$^+$ (0) &  & 0.17 \Correction{$\left(0.01\atop0.01\right)$} & $-0.11$ \Correction{$\left(0.01\atop0.02\right)$} \\
2233.6 & $^{32}$S$(n,d)^{31}$P & $\dfrac{5}{2}^+$ (2233.6) & $\dfrac{1}{2}^+$ (0) & & & \\
\hline
2775.6 & $^{32}$S$(n,n')^{32}$S & 3$^-$ (5006.2) & 2$^+$ (2230.6) & This work & $-0.09$ \Correction{$\left(0.02\atop0.01\right)$} & * \\

\bottomrule

\end{tabular}

\end{center}

\begin{multicols}{2}

\section{Conclusion}
The main goal of this work was to study the characteristics of $\gamma$ radiation emitted by atomic nuclei from the reactions induced by $14.1$~MeV neutrons. In the course of its implementation, measurements of the yields and angular distributions of $\gamma$-quanta were performed with two different experimental setups, for samples containing natural oxygen, phosphorus and sulfur.
The joint use of two different detector systems, as well as the improvement of the data processing procedures allowed us to significantly increase the reliability of the experimental data.

Comparison of the experimentally obtained yields of $\gamma$-rays with the data of other published works and with the data of the nuclear data library shows, on the whole, a fairly good agreement between them. At the same time, quite significant discrepancies are observed for several individual $\gamma$-transitions. The agreement between the experimental yields of $\gamma$-quanta and those theoretically obtained using the TALYS 1.96 program for most of the observed transitions is quite good, with the exception of the reactions $(n,p)$ $(n,d)$ and $(n,\alpha)$, which is obviously due to the fact that nuclei with insufficiently known parameters of their optical potentials are formed in these reactions.

Despite the long study of neutron-nuclear reactions, there is still much work to be done in this field. Data on the yields and emission cross-sections of $\gamma$-quanta at certain energies are needed for the currently developed methods for rapid elemental analysis of various substances using $D-D$ and $D-T$ neutron generators as neutron sources. 

The use of portable neutron generators for fast and high-quality elemental analysis requires the use of large samples and long measurement times due to the relatively low intensity of the neutron flux from these devices. Therefore, the development and application of methods for correcting the obtained data for the scattering and absorption of $\gamma$-quanta and neutrons in large samples is necessary, and the results of this work show that this is quite feasible.

\section{Acknowledgements}
The authors are grateful to VNIIA (Moscow, Russia) for providing a portable neutron generator ING-27. We thank D.~N.~Borisov and S.~I.~Negovelov for their help in preparing the experiments.

\appendix
\section*{APPENDIX A: Optical model formalism}
The default optical model (OM) parameter set used in TALYS are the local and global parameterisations of Koning and Delaroche.

The phenomenological optical potential $U$ for nucleon-nucleus scattering is usually  defined as
\begin{multline}
U(r, E) = - \mathcal{V}_{V}(r, E) - i\mathcal{W}_{V}(r, E) - i\mathcal{W}_{D}(r, E) + \\+(\Vec{l} \cdot \Vec{\sigma})(\mathcal{V}_{SO}(r, E) + i\mathcal{W}_{SO}(r, E)) + \mathcal{V}_{C}(r),
\label{OptPotential}
\end{multline}
where $V_{V,SO}$ and $W_{V,D,SO}$ are the real and imaginary components of the volume-central $V$, surface-central $D$ and spin-orbit $SO$ potentials, respectively, $l$ and $\sigma$ are spin and angular momentum operators.  All components are separated in $E$-dependent well depths, $V_{V}$, $W_{V}$ ,$W_{D}$, $V_{SO}$, and $W_{SO}$, and energy-independent radial parts $f$, namely

\begin{equation*}
    \mathcal{V}_{V}(r, E) = V_{V}(E)f(r, R_{V}, a_{V}),
\end{equation*}
\begin{equation*}
    \mathcal{W}_{V}(r, E) = W_{V}(E)f(r, R_{V}, a_{V}),
\end{equation*}
\begin{equation*}
    \mathcal{W}_{D}(r, E) = -4a_{D}W_{D}(E)\dfrac{d}{dr}f(r, R_{D}, a_{D}),
\end{equation*}
\begin{equation*}
    \mathcal{V}_{SO}(r, E) = V_{SO}(E) \lambdabar ^{2}_{\pi}\dfrac{1}{r}\dfrac{d}{dr}f(r, R_{SO}, a_{SO}),
\end{equation*}
\begin{equation}
    \mathcal{W}_{SO}(r, E) = W_{SO}(E) \lambdabar ^{2}_{\pi}\dfrac{1}{r}\dfrac{d}{dr}f(r, R_{SO}, a_{SO}),
\end{equation}
\label{Opt1}
where geometry parameters are the radius $R_{i} = r_{i}A^{1/3}$ and diffuseness parameter $a_{i}$ determines form-factor $f(r, R_{i}, a_{i})$.

All components of the potential depend on $(E-E_{F})$, where $E_{F}$, the Fermi energy for neutron or proton in MeV is defined as the energy halfway between the last occupied and the first unoccupied shell of the nucleus
\begin{equation*}
    E^{n}_{F} = -\frac{1}{2}[S_{n}(Z,N) + S_{n}(Z,N+1)],
\end{equation*}
\begin{equation}
    E^{p}_{F} = -\frac{1}{2}[S_{p}(Z,N) + S_{p}(Z+1,N)].
\end{equation}

TALYS uses OMP parametrisation given below:
\begin{equation*}
    V_{V}(E)= v_{1}[1-v_{2}(E-E_{F})+v_{3}(E-E_{F})^{2}-v_{4}(E-E_{F})^{3}],
\end{equation*}
\begin{equation*}
    W_{V}(E)= w_{1} \frac{(E-E_{F})^{2}}{(E-E_{F})^2 + (w_{2})^{2}},
    \end{equation*}
\begin{equation*}
    W_{D}(E)=d_{1}\frac{(E-E_{F})^{2}}{(E-E_{F})^{2} + (d_{3})^{2}}\exp[-d_{2}(E-E_{F})],
\end{equation*}
\begin{equation*}
    V_{SO}(E)=v_{so1}\exp[-v_{so2}(E-E_{F}],
\end{equation*}
\begin{equation}
    W_{SO}(E)=w_{so1}\frac{(E-E_{F})^{2}}{(E-E_{F})^{2} + (w_{so2})^{2}}
\label{Opt3}
\end{equation}

Parameters $r_{V}$, $a_{V}$, $r_{D}$, $a_{D}$, $r_{SO}$, $a_{SO}$, $r_{C}$ do not depend on energy. This parametrisation is valid for incident energy $E$ from 1 keV up to 200 MeV.

For nuclei discussed in this paper, predefined OM parameters and deformations available for $^{31}$P and $^{32}$S. It is important to notice that sulfur is a quite attractive object for optical model testing \cite{Wu2014,AlOhali2012}. For $^{16}$O they were calculated using Koning parametrisation. In Table~\ref{Tab:OMP} OM parameters are listed according to Eq.~(\ref{Opt3}).

\begin{center}
\tabcaption{\label{Tab:OMP} Optical model parameters.}
\footnotesize
\begin{tabular}{ l  l  l  l }
\toprule
Nuclide	& $^{16}$O &	$^{31}$P &	$^{32}$S \\ 
\hline
$v_1$ & $ 58.916 $ & $ 57.8 $ & $ 59.5$ \\ 
$v_2$  & $ 7.2\times 10^{-3}$ & $ 7.2\times 10^{-3}$ & $ 7.2\times 10^{-3}$ \\
$v_3$  & $ 1.962\times 10^{-5}$  & $ 1.9\times 10^{-5}$  & $ 1.9\times 10^{-5}$ \\
$v_4$  & $ 7.1\times10^{-9}$ & $ 0 $ & $ 0 $\\
$w_1$  & $ 12.462 $ & $ 12.4 $ & $ 12.6$ \\
$w_2$  & $ 74.822 $ & $ 76 $ & $ 75 $\\
$d_1$  & $ 16 $ & $ 15.4 $ & $ 15.6 $\\
$d_2$  & $ 0.0218 $ & $ 0.0214 $ & $ 0.0215 $\\
$d_3$  & $ 11.5 $ & $ 11.5 $ & $ 11 $\\
$v_{so1}$  & $ 5.97 $ & $ 6 $ & $ 6 $\\
$v_{so2}$  & $ 0.004 $ & $ 0.004 $ & $ 0.004 $\\
$w_{so1}$  & $ -3.1 $ & $ -3.1 $ & $ -3.1 $\\
$w_{so2}$  & $ 160 $ & $ 160 $ & $ 160 $\\
Approach & DWBA & DWBA & VIB \\
\bottomrule

\end{tabular}
\end{center}

\section*{\Correction{APPENDIX B: Angular distribution of $\gamma$-quanta from ENDF}}

\Correction{According to the ENDF-6 format manual describing the Evaluated Nuclear Data Files ENDF/B-VI, ENDF/B-VII and ENDF/B-VIII (eq. (14.1) in \cite{ENDF-manual}), the angular distribution of $\gamma$-radiation in ENDF format is described in terms of $p_k(\theta,E)$, i.e. probability of emission of $\gamma$-quantum with number $k$ emitted as a result of nuclear reaction induced by particle with energy $E$.}

\Correction{\begin{equation}
p_k(\theta,E)=\frac{2\pi}{\sigma_k^\gamma (E)}\frac{d\sigma_k^\gamma (E)}{d\Omega},
\label{eq:ENDF-prob}
\end{equation}}

\Correction{on the other hand, in ENDF file $p_k(\theta,E)$ is presented in terms of Legendre series expansion:}

\Correction{\begin{equation}
p_k(\theta,E)=\sum_{l=0}^{2J} \frac{2l+1}{2} b_l^k(E)P_l(\cos\theta),
\label{eq:ENDF-legendre}
\end{equation}}
\Correction{where $J$ is multipolarity of the $\gamma$-quantum. Formula (\ref{eq:legendre}) can be represented as following expression:}

\Correction{\begin{equation}
\label{eq:formula5}
\frac{4\pi}{\sigma_k^\gamma (E)}\frac{d\sigma_k^\gamma (E)}{d\Omega}=1+\sum_{l=2,4...}^{2J}a_l^k(E)P_l(\cos\theta),
\end{equation}}
\Correction{so, eq. (\ref{eq:ENDF-legendre}) and (\ref{eq:formula5}) differ only in normalization coefficient, and we converted $b_l^k$ values to $a_l^k$ using the following formula:}

\Correction{\begin{equation}
a_l^k=(2l+1) b_l^k.
\end{equation}}

\Correction{The $b_l^k(E)$ coefficients were extracted from the ENDF-B/VIII.0 database, file "n\_0825\_8-O-16", MF=14, MT=4 rows 36449-36452 for 6128.9~keV transition in $^{16}$O and MF=14, MT=107, rows 38118-38121 for 3684.5~keV transition in $^{13}$C. We have interpolated values of $b_l^k$ corresponding for two projectile energies to obtain value for 14.1 MeV neutrons.}

\end{multicols}

\vspace{15mm}

\vspace{-1mm}
\centerline{\rule{80mm}{0.1pt}}
\vspace{2mm}

\begin{multicols}{2}

\end{multicols}

\clearpage
\end{CJK*}
\end{document}